\documentclass[pra,
twocolumn,
superscriptaddress,showpacs,floatfix,nofootinbib]{revtex4}
\usepackage{epsfig,amsmath,amsfonts,amssymb}
\usepackage{graphicx}
\usepackage[utf8]{inputenc}
\usepackage{boxedminipage}
\usepackage{epsfig,psfrag}
\usepackage{amsmath}
\usepackage{mathbbol, amsfonts}
\usepackage{lscape}
\usepackage{fancybox}
\usepackage{amsfonts,amssymb}
\usepackage{color}
\newcommand{\beq}{\begin{equation}}
\newcommand{\eeq}{\end{equation}}

\newcommand{\beqa}{\begin{eqnarray}}
\newcommand{\eeqa}{\end{eqnarray}}
\def\ra{\rangle}
\def\la{\langle}

\usepackage{wrapfig}

\newcommand{\be}{\begin{equation}}
\newcommand{\ee}{\end{equation}}

\begin{document}

\title{Discriminating between the von Neumann and Lüders reduction rule}
\author{G. C. Hegerfeldt}
\affiliation{Max Planck Institute for the Physics of Complex Systems,
N\"othnitzer Str. 38 D-01187 Dresden, Germany}
\affiliation{Institut für Theoretische Physik, Universität Göttingen,
Friedrich-Hund-Platz 1, D-37077 Göttingen, Germany}

\author{R. Sala Mayato}
\affiliation{Max Planck Institute for the Physics of Complex Systems,
N\"othnitzer Str. 38 D-01187 Dresden, Germany}
\affiliation{Departamento de F\'\i sica Fundamental II and IUdEA,
Universidad de La Laguna, La Laguna, 38204, S/C de Tenerife, Spain}

\begin{abstract}

Given an ensemble of systems in an unknown state, as well as 
 an observable $\hat A$ and a physical  apparatus which performs a measurement
 of $\hat A$ on the ensemble,  whose detailed working is unknown
 ('black box'),  how can one test whether the  Lüders or von Neumann
 reduction rule applies?

\end{abstract}
\maketitle

\section{Introduction}

In his ground-breaking book \cite{vN} of 1932, von Neumann
investigated the quantum mechanical measurement problem and formulated
a rule how to obtain the state of an ensemble of physical systems
after a measurement.  This rule was later substantially modified by Lüders 
\cite{L}. It is the Lüders reduction or projection rule that nowadays is
mostly used. The Lüders rule states that after a selective measurement
\cite{selective} 
of an observable $\hat A$ with discrete eigenvalues
the subensemble of systems with the measurement result $a_i$
is in the (non-normalized, pure) state $\hat P_i |\psi\rangle$, where
$\hat P_i$ is 
the (possibly multi-dimensional) projection operator onto the
eigenspace of the eigenvalue $a_i$ 
and where $|\psi\ra$ the state prior to the measurement. For an
initial density matrix $\hat\rho$ one obtains $\hat P_i \hat\rho \hat
P_i$ \cite{remark}.

The von Neumann reduction rule assumes  that in the case of degeneracy
one
measures a refinement $\hat A'$ of $\hat A$,   which
commutes with $\hat A$ and which has only non-degenerate discrete
eigenvalues \cite{vN} and thus lifts the degeneracy. Usually this
results from consecutive measurements (for examples
cf. e.g. \cite{Rafa} and Eq. (\ref{basis2}) of the Appendix below).  
 Then $\hat A$ is a
function of $\hat A'$, $\hat A = f(\hat A')$, say, and from a
measurement result $a'$ of $\hat A'$ for an individual system one
obtains the corresponding result $a=f(a')$ for $\hat A$.
As a generalization  we introduce here the notion
of a {\em partial} von Neumann type measurement  which can also arise
from consecutive measurements (cf. Appendix). One can choose
a refinement which lifts the degeneracy of $\hat A$ only
partially. Then still  $\hat A = f(\hat A')$, but $\hat A'$ may have some
degenerate eigenvalues. Then a
partial von Neumann type measurement is obtained by performing a Lüders type
measurement of this observable $\hat A'$.
There are recent  and important investigations of  the state after more
general measurements \cite{Beltrametti,recent}, but we restrict
ourselves to the above reduction rules.

In this paper we propose a simple three-step procedure, based on selective
measurements, to test whether one deals with a Lüders type measurement of
an observable  $\hat A$ or
not. We illustrate this for a particular measurement result, $a_1$
say. After the measurement  the subensemble of systems with the
 result $a_1$ is selected and denoted by ${\cal E}_1$.
Then a refinement of $\hat A$, denoted by $\hat \sigma$, with discrete
non-degenerate eigenvalues, is measured for each system of ${\cal
  E}_1$.  Then on ${\cal E}_1$ one again measures $\hat A$  by means
of the unknown apparatus and then again $\hat \sigma$. If for any
system of ${\cal  E}_{1}$ the  result of the second measurement of
$\hat \sigma$  differs from the first, one does not have a Lüders type
measurement. If  the  results are the same for each system  one chooses
another, 
particular,  refinement $\hat \sigma'$ of $\hat A$ with non-degenerate
eigenvalues and which 
does not commute with  $\hat \sigma$ (i.e. only with $\hat A$). Then
one proceeds as before, with $\hat \sigma'$ instead of $\hat \sigma$.
 But now it turns out that one has a Lüders type measurement (on ${\cal E}_1$) if
 and only if for  each system the two results  of the $\hat \sigma'$
 measurements  are the same. 

The plan of the paper is as follows. In Section \ref{two} we show how
the procedure works 
in the simple case of an observable with twofold degeneracy. In
Section \ref{n} the general case is treated. In Section \ref{disc} the
results are discussed. In the Appendix we give
examples of Lüders, von Neumann, and partial von Neumann type
measurements.

\section{Testing the twofold degenerate case}\label{two}

For greater transparency the
procedure will first be explained  for the example of the Appendix
with two spins, 
$\hat A=\sigma_{1z} + \sigma_{2z}$.  The 
eigenvalue $a_1 = 0$ of $ \hat A $ is twofold degenerate. An as yet
unknown apparatus performs a measurement of $\hat A$ on an
ensemble $\cal E$. The apparatus can be assumed to perform a
measurement of an observable $\hat A'$, which is a possible trivial or
nontrivial refinement of $\hat A$. We  say that $\hat A'$ is
associated to the apparatus. 

 We assume that the result $a_1 = 0$ is found on  a subensemble 
${\cal E}_1$ of systems. 
In the two-dimensional eigenspace of the eigenvalue $a_1$ of $\hat A$
the as yet unknown observable $\hat A'$,  which commutes with $\hat
A$, either has two non-degenerate eigenvalues or  a single twofold
degenerate eigenvalue. In the former case the apparatus performs a von Neumann
measurement and in the latter a Lüders measurement.  

Now we choose a refinement $\hat\sigma$  of $\hat A$ with
non-degenerate eigenvalues. As an example, we 
take it to be diagonal in the basis $|++\ra $, $|+-\ra $, $|-+\ra
$, and  $|--\ra $, e.g. 
\be
\hat \sigma = \sum_{ij=\pm }(2i+j)|ij\ra\la ij|~.
\ee
In the two-dimensional eigenspace of $\hat A$ for $a_1$ the
eigenvalues and eigenvectors of $\hat \sigma $ are $s_1=1$ with
$|s_1\ra = |+-\ra$ and $s_2=-1$ with $|s_2\ra= |-+\ra$.
It may happen that, inadvertently and at this stage unknown to us, the
chosen  $\hat \sigma$ and the unknown operator $\hat A'$ associated
with the apparatus are jointly diagonal and commuting. This will bring
a complication and will later require an additional step in the procedure. 
Now we proceed as follows.

(i) First a  measurement of $\hat \sigma$ is performed on the
subensemble ${\cal E}_1$. Since $\hat \sigma$ has only non-degenerate 
eigenvalues there is no difference between  a von Neumann and a Lüders
measurement of $\hat \sigma$ and the possible results are  $s_1 $ and $s_2$. 
If $s_1$ is found the subensemble of corresponding systems in ${\cal
  E}_1$ is denoted by ${\cal E}_{11}$, and after this measurement it is in
the pure state $|s_1\ra$. Similarly for  $s_2$. 

(ii) Now one lets the apparatus measure $\hat A$ on the systems
of ${\cal   E}_{1}$. Of course, the value found
is again $a_1$ for each system. If the apparatus performs a Lüders
measurement the state $|s_i\ra$ of subensemble ${\cal E}_{1i}$ is definitely not
changed, while  for a von Neumann measurement a change of the state
$|s_i\ra$ may or may not occur.

(iii) After this one again
measures $\hat \sigma$ on ${\cal E}_{1}$. If  one finds both $s_1$ and
$s_2$ for systems in  ${\cal   E}_{11}$, then  
the state $|s_1\ra$ has been changed and one concludes
that the apparatus  performs a von Neumann measurement. Similarly for 
${\cal E}_{12}$.   

If, on the other hand,  one finds only $s_1$ on ${\cal E}_{11}$ 
this  means that one of the  projection operators in the decomposition of
$\hat A'$  leaves $|s_1\ra$ invariant and that  $|s_1\ra$ is an
eigenvector of $\hat A'$. But then the orthogonal vector $|s_2\ra$ is
also an eigenvector of $\hat A'$. Thus
  $\hat A'$
and  $\hat \sigma$ are diagonal in the same basis and commute.
%
To find out if the above unknown  projection operator of $\hat A'$ is
two-dimensional or not  one now chooses another operator, $\hat
\sigma'$ say, which does not commute with $\hat \sigma$, e.g.    
\be
\hat \sigma' =  
\sigma_{1z} + \sigma_{2z}+   (\vec \sigma_1+\vec\sigma_2)^2~.
\ee
The relevant eigenvalues are $s_{1}'= 4$ and $s_{2}'= 0$  with
eigenvectors $|s_{1,2}'\ra = |\phi_{+,-} \ra$ from Eq. (\ref{vN1}), and
they are not orthogonal to $|s_{1}\ra$, with $\la s_1|s_{1,2}'\ra\neq 0$. Now one
proceeds as before for $\hat \sigma$. One first measures $\hat
\sigma'$ on the systems of 
 ${\cal E}_{11}$, which is in the state $|s_1\ra$,  and denotes
by ${\cal E}_{11}'$ the subensemble of systems for which the value
$s_1'$ has been found. Then, on
${\cal E}_{11}'$,  which is in the state $|s_1'\ra$, one lets the
apparatus perform a measurement of $\hat A$. This measurement again yields the
value $a_1 $, but it may or may not have changed the state of ${\cal
  E}_{11}'$, depending on whether the apparatus performs a von Neumann
or Lüders measurement. Then again $\hat \sigma'$ is measured on ${\cal
  E}_{11}'$. If both $s_1'$ and $s_2'$ appear  as measurement results
the state has been changed and therefore the 
apparatus performs a von Neumann measurement.
If only the value $s_1'$ appears then $|s_1'\ra$ is
an eigenstate of ${\hat A}'$, as is
$|s_1\ra$, from the above argument. From the non-orthogonality of
$|s_1\ra$ and  $|s_1'\ra$  it follows that the
corresponding eigenspace of ${\hat A}'$ is two-dimensional. Hence in
this case the apparatus performs a Lüders measurement.

\section{The general test}\label{n}
In this section we describe the test procedure for Lüders vs. von
Neumann for a general observable $\hat A$ with discrete, possibly
degenerate, eigenvalues $a_k$. Corresponding orthogonal eigenvectors
are denoted by $|a_{k}^\alpha\ra$, $\alpha = 1, \cdots, n_k$ so that
the degeneracy is $n_k$. Then 
\be
\begin{split}
\hat A_k &= \sum_{k,\alpha}a_k |a_{k}^\alpha\ra\la a_k^\alpha|\\
         &\equiv \sum_k a_k  \hat P_k
\end{split}
\ee
where $\hat P_k$ is the  projection operator onto the eigenspace of $a_k$.
According to the
Lüders rule the subensemble with the measurement result $a_k$ is
described by   
\be
 \hat P_k |\psi\ra
\ee
in case of a pure initial state $|\psi\ra$, and by
\be\label{Li}
\hat P_k \hat \rho \hat P_k
\ee
in case of a mixed initial state $\hat \rho$. Both the norm squared
and the trace give the probability of finding the value
$a_k$. The complete ensemble is, directly after the measurement,
described by the normalized density matrix 
\be\label{Lred}
  \sum_k \hat P_k \hat \rho \hat P_k\,.
\ee

Now consider a refinement $\hat A'$ of $\hat A$ which partially lifts
the degeneracy of  $\hat A$. Then  $\hat A'$  is of the form
\beqa\label{partial}
A' &=& \sum_{k\beta} a'_{k\beta} \hat P_k^\beta\label{A'}\\
\hat P_k &=& \sum_{\beta=1}^{m_k}  \hat P_k^\beta\label{alpha}
\eeqa
where the orthogonal  projection operators $\hat P_{k}^\beta, ~\beta =
1\cdots m_k$,  
 are partial sums of $|a_{k}^\alpha\ra\la a_{k}^\alpha|$ for
 fixed $k$. Then $\hat A$ is a function of $\hat A'$, $\hat A=f(\hat
 A')$, and $f(a_{k\beta}')= a_k$. A partial von Neumann measurement of
 $\hat A$ is obtained by a Lüders measurement of $\hat A'$ where the
 apparatus is so programmed that its output is $f(a_{k\beta}')$ instead
 of $a_{k\beta}'$.  After the measurement the subensemble for
 which the output is  $f(a_{k\beta}')=a_k$ is
 described instead of by Eq.(\ref{Li}) by the density  matrix  
\be \label{pNk}
  \sum_\beta \hat P_k^\beta\,\hat \rho\, \hat P_k^\beta
\ee
and the complete ensemble by their sum  over $k$. Note that
if $m_k=1$ for all $k$, i.e. $\hat P_k^\beta \equiv \hat P_k^1=\hat P_k$,
then one has a Lüders measurement, and if all $\hat 
P_k^\beta$ are one-dimensional  projection operators one has a usual
(i.e. not a partial) von Neumann measurement.

We now describe the test procedure Lüders vs. von Neumann for the
general case and consider an ensemble ${\cal E}$ of systems with
initial density matrix $\hat\rho$. As before we denote by $\hat A'$ the
observable associated to the unknown apparatus and consider the subensemble 
${\cal E}_1$ of systems for which 
an eigenvalue $a_1$ of $\hat A$ has been found as the measurement result. The
eigenvalue is $n_1$ fold degenerate. The subensemble ${\cal E}_1$ is
described by the density  matrix $\hat \rho_1$, with
\be
\hat \rho_1 = \sum_{\beta=1}^{m_1} \hat P^\beta_{1}\,\hat \rho\,
\hat P_1^\beta~,
\ee
where the $\hat P_k^\beta$ are the unknown  projection operators on
eigenspaces of $\hat A'$.  

Let $\hat \sigma$ be an observable  commuting with $\hat A$ and with discrete
non-degenerate eigenvalues. The previous steps can now be adapted as follows.

(i) First one measures $\hat \sigma$ on the systems of ${\cal E}_1$. The
eigenvalues of $\hat \sigma$ in the $a_1$ eigenspace  are denoted
 by $s_1, \cdots, s_{n_1}$, with eigenstates $|s_i\ra$. The
 subensemble of systems for which the result $s_i$ is found in the
 measurement will be
 denoted by ${\cal E}_{1i}$. It  can be described  by the  pure
 state $|s_i\ra$.  

(ii)  Now one lets the apparatus perform a measurement of $\hat A $ on
the systems of ${\cal E}_{1}$. The result is of course again $a_1$ and after
the measurement the density matrix of the subensemble ${\cal E}_{1i}$
is proportional to  
 \be \label{iia}
  \sum_{\beta=1}^{m_1}
 \hat P^\beta_{1}\,|s_i\ra \la s_i|\,\hat P^\beta_{1}~.
\ee
If $m_1=1$, i.e. if the apparatus performs a Lüders measurement in the
$a_1$ eigenspace, this is the pure state $|s_i\ra\la s_i|$. Otherwise
it is a mixed state. 

(iii) After this, one again measures $\hat \sigma$ on ${\cal E}_{1}$. If for
a system of a subensemble ${\cal E}_{1i}$ this second measurement
of $\hat \sigma$ gives a result different from $s_i$ one
concludes from step (ii) that the apparatus  has changed the state
$|s_i\ra$ and thus has not performed a Lüders measurement, but rather
a (possibly partial) von Neumann measurement. 

If, however, the result is always $s_i$ for each
${\cal E}_{1i}$  then ${\cal E}_{1i}$ remains in its state $|s_i\ra$
after the measurement of $\hat A$ by the apparatus
and hence this state is an eigenvector of $\hat A'$. (If all $s_i$
appear as measurement results this implies that $\hat A'$ and $\hat
\sigma$ happen to commute on the $a_1$ eigenspace.)

In this case  one chooses an additional observable  $\hat
\sigma '$, with non-degenerate eigenvalues  and which commutes  with $\hat
A$ but not with $\hat \sigma$ and which has the following special
property. In the  $a_1$ eigenspace  the 
 eigenvalues and eigenvectors of $\hat \sigma'$ are denoted by  $s_i'$
 and $|s_i'\ra$. The latter are connected to the eigenvectors
 $|s_j\ra$ of $\hat \sigma$ by a unitary transformation, and one
 chooses $\hat \sigma'$ in such a way that  one has
\be
\begin{split} \label{s'}
 |s_1\ra    = \sum_{i=1}^{n_1} \gamma_{i } |s_i'\ra ~~~~~~{\rm with}~~
 \gamma_{i}\neq  0~~~{\rm for~ all}~i
\end{split}
\ee
where $s_1$ is assumed to have occurred as a result in the measurement
of $\hat \sigma$. Such a $\hat \sigma'$ can always be found, and
Eq. (\ref{s'}) is the key to distinguishing both types of measurements.
On the systems of the subensemble ${\cal  E}_{11}$
(which in this case has remained in the  state  $|s_1\ra$) one then
performs,  with $\hat \sigma$ replaced by $\hat \sigma '$, the steps
(i)-(iii) .  Since the transition probability $| \la s_1|s_j'\ra|^2 =
|\gamma_j|^2\neq 0$, all eigenvalues $s_i'$ appear as measurement results
in the first measurement of $\hat \sigma'$, and the associated
subensembles ${\cal E}_{1i}'$ are in the state $|s_i'\ra$. Then, as the
second step, one lets
the apparatus perform a measurement of $\hat A$. In the third step
$\hat \sigma'$ is 
measured again on the systems of the subensemble ${\cal  E}_{11}$. If
for any system of ${\cal  E}_{11}$ the  result of the second
measurement of $\hat \sigma'$  differs from the first then the state has
been changed by the apparatus  and one has a (possibly partial)
von Neumann measurement. 

Otherwise, if for all systems of ${\cal  E}_{11}$  the  result of the second
measurement of $\hat \sigma'$ is the same as in the first, then 
 the states $|s_i'\ra$ are not changed and hence
are eigenvectors of $\hat A'$, as is $|s_1\ra$. Then all vectors in 
Eq. (\ref{s'}) are eigenvectors of $\hat A'$. But  this can only 
happen if they belong to the same eigenvalue since $|s_1\ra$ is not
orthogonal to any $|s_i'\ra$. This implies that
the $a_1$ eigenspace of $\hat A$ is also an eigenspace of $\hat A'$
and hence the apparatus performs a Lüders measurement of $a_1$ if for
each system of the subensemble ${\cal  E}_{11}$ the
results of the first and second measurement of $\hat \sigma'$
are the same. 

\section{Discussion} \label{disc}

In this paper the two forms of the reduction rule due to von Neumann
and Lüders, also known as the projection postulate,  have been
discussed. The original 
formulation of von Neumann starts with an observable with discrete,
possibly degenerate, eigenvalues, but then goes over to a refinement
with non-degenerate eigenvalues, thus lifting the degeneracy. The
projection operators are then one-dimensional and project onto the
individual non-degenerate 
eigenvectors. Lüders, on the other hand, does not lift the degeneracy
but uses projections onto eigenspaces of the original observable. The
dimension of these eigenspaces are given by the degeneracy of the
observable under consideration. In
this paper we have also introduced an additional, sort of  
intermediary, reduction rule for which a refinement of the observable
is used which lifts the degeneracy only partially and which may
retain some degeneracy. We call the associated measurement  a
partial von Neumann measurement.

It has been shown here that all three forms of the
reduction rule may appear quite naturally, depending on the
realization of a particular measurement apparatus. Therefore all three forms
have their own legitimacy, and one can not say that one is better than the
other. Their applicability depends on the circumstances, i.e. the
details of the measurement apparatus.

The main investigation of this paper focused on the following
question. If a measurement apparatus for an observable is only known
to obey one of the 
forms of the reduction rule of von Neumann and Lüders, but otherwise the 
details of the  apparatus  are not known how
can one check whether the reduction has occurred by the Lüders rule or
not? To this end we have proposed and studied a three-step procedure
based on measurements of an auxiliary observable. The outcome of  the
latter measurements indicates the type of reduction. 

It would be interesting if one could carry this investigation over to
the more general types of measurements characterized in
Ref. \cite{Beltrametti}. 

\section*{Acknowledgements}
 We thank J.G. Muga for discussions. 
We also acknowledge the kind hospitality of the Max Planck 
Institute for Complex Systems in Dresden, and   
funding of R. Sala Mayato  by Ministerio de 
Ciencia e Innovaci\'on (Grant No. FIS2010-19998).

\begin{appendix}
\section{Examples  for different measurement schemes}

We consider an ensemble consisting of systems, each with two
independent spins, 
$\vec\sigma_1$ and $\vec\sigma_2$, with $\vec\sigma$ the Pauli
matrices. 
For the $z$ component of the total spin,
$\sigma_{{\rm tot},z}=\sigma_{1z} + \sigma_{2z}$ the possible
measurement results are 2, 0, 0, -2, with corresponding eigenvectors
$|++\rangle$, $|+-\rangle$, $|-+\rangle$, and $|--\rangle$.  

Now, in case of a Lüders type measurement, if the initial
state of the ensemble is a pure state $|\psi\rangle$ then 
 after the measurement the respective subensembles are given  by the
 pure states $\hat  
P_i|\psi\rangle$, $i=1,~0,~-1$. Here 
\be \label{proj1}
\hat P_1 =\hat P_{|++\rangle},~\hat P_0=\{\hat P_{|+-\rangle} + \hat
P_{|-+\rangle} \},~\hat P_{-1}=\hat P_{|--\rangle}
\ee
 where $\hat
P_{|\phi\rangle}\equiv |\phi\rangle\langle\phi| $.
 The total
ensemble is then described after this measurement by a density matrix
given by $\sum_i\hat P_i |\psi\ra\la\psi|\hat P_i$.
Similarly for an initial density matrix instead of a pure state.

Following von Neumann, instead of $\sigma_{{\rm tot},z}$ one can 
measure a refinement of $\sigma_{{\rm tot},z}$ with non-degenerate
eigenvalues, e.g. the observable $\hat  
A'\equiv  \sigma_{{\rm tot},z} +  (\vec
\sigma_{\rm tot})^2$, which lifts the degeneracy of $ \sigma_{{\rm
    tot},z}$. From this one obtains an indirect measurement of 
$\sigma_{{\rm tot},z}$ as follows. The  
eigenvectors of $\hat A' $ are 
\be \label{vN1}
\begin{split}
|++\rangle\,,~~~~|\phi_+\rangle\equiv\{|+-\rangle
+|-+\rangle\}/\sqrt{2}\,,\\
|--\rangle\,,~~~~|\phi_-\rangle\equiv\{|+-\rangle -
|-+\rangle\}/\sqrt{2}\,,
\end{split}
\ee
 with respective eigenvalues $a'= 6$, 4, 2 and 0. If 
one defines the function $f(x) = -\frac{8}{3}x+ x^2 - \frac{1}{12}x^3$
then $f(6)= 2$, $f(4)= 0$, $f(2)=-2$ and $f(0)=0$, and
$f(\hat A') = \sigma_{{\rm tot},z}$.
This is either checked directly by insertion of $\hat A'$ or by applying
$f(\hat A')$ to the eigenvectors of $\hat A'$. Therefore, if the
result of an $\hat A'$ measurement on a system is $a'$, then one knows that
$\sigma_{{\rm tot},z}$ has the values $f(a')$. In contrast to the
previous Lüders measurement, now the subensemble with
the result 0 for $\sigma_{{\rm tot},z}$ is in a mixed state, given by
the density matrix  
\be
\begin{split}
\hat P_{|\phi_+\rangle}|\psi\rangle \langle \psi|\hat
P_{|\phi_+\rangle} + \hat P_{|\phi_-\rangle}|\psi\rangle \langle
\psi|\hat P_{|\phi_-\rangle}~.
\end{split}
\ee
 The complete ensemble has now the density matrix 
\be \label{basis1}
\begin{split}
&\hat P_{|++\rangle}|\psi\rangle \langle \psi|\hat
P_{|++\rangle} 
+ \hat P_{|\phi_+\rangle}|\psi\rangle \langle \psi|\hat
P_{|\phi_+\rangle}\\ 
&+ \hat P_{|\phi_-\rangle}|\psi\rangle \langle
\psi|\hat P_{|\phi_-\rangle}
 + \hat P_{|--\rangle}|\psi\rangle \langle \psi|\hat
P_{|--\rangle}~.
\end{split}
\ee

For consecutive measurements von Neumann type measurements appear quite
naturally.  If         
the two spins are spatially sufficiently separated one can
measure them  individually, e.g. first a Lüders measurement of
$\sigma_{1z}$ and then immediately afterwards of $\sigma_{2z}$
\cite{Rafa}. This also provides a measurement of $\sigma_{{\rm tot},z}$. 
In this case, the possible individual measurement
results are $++,~ +-,~ -+,~ --$, and after the measurement the
corresponding subensembles are 
obviously in the states $|++\rangle$, $\cdots$, $|--\rangle$. If the
initial state of the ensemble is a pure state $|\psi\rangle$ then
after the measurement its state is given by the density matrix 
\be \label{basis2}
\sum_{i,j=\pm} \hat P_{|ij\rangle}|\psi\rangle\langle \psi |\hat
P_{|ij\rangle} 
\ee
and similarly for an initial density matrix. It is apparent that this
consecutive measurement amounts to a particular von Neumann
measurement, but with a resulting density matrix which differs from
the previous one in Eq. (\ref{basis1}). The measurement is equivalent
to a separate measurement of 
the projection operators $\hat P_{|ij\rangle}$ or, equivalently, of an
observable of the form $\hat A'=\sum a'_{ij}\hat P_{|ij\ra}$ with
pairwise different 
$a'_{ij}$'s. This observable $\hat A'$ can be taken to be associated
with the apparatus measuring the individual spins separately.

We now consider an
ensemble consisting of systems each with three independent spins, 
$\vec\sigma_1$, $\vec\sigma_2$, and $\vec\sigma_3$, and the observable
$\hat A = \sigma_{1z} + \sigma_{2z}$.   
One can now  consecutively measure $\sigma_{1z}$ and $\sigma_{2z}$, as
before. Then with three independent spins, this gives a partial von
Neumann measurement of $\sigma_{1z} + \sigma_{2z}$. If one
measured all three 
spins consecutively this would resolve the degeneracy and lead to an
ordinary von Neumann measurement.

\end{appendix}


\begin{thebibliography}{99}
\bibitem{selective}
A selective measurement is a measurement procedure in which the result
for each individual system is recorded.
\end{thebibliography}
\end{document}